\documentclass[12pt,preprint]{aastex}

\newcommand{\Msol}{$M_\odot$}
\newcommand{\Rsol}{$R_\odot$}
\newcommand{\DC}{$\delta$\,Cephei}
\newcommand{\kms}{km\,s$^{-1}$}
\newcommand{\ms}{m\,s$^{-1}$}
\newcommand{\hermes}{{\it Hermes}}

\slugcomment{To be published in the Astrophysical Journal}

\shorttitle{Discovery of $\delta$\,Cep's Secret Companion}
\shortauthors{Anderson, R.I. et al.}

\begin{document}

\title{Revealing $\delta$\,Cephei's Secret Companion and Intriguing Past}

\author{R.I. Anderson\altaffilmark{1,2,4}, J. Sahlmann\altaffilmark{3}, 
B. Holl\altaffilmark{2}, L. Eyer\altaffilmark{2}, L.
Palaversa\altaffilmark{2}, N. Mowlavi\altaffilmark{2}, M.
S\"uveges\altaffilmark{2}, M. Roelens\altaffilmark{2}}
\affil{$^1$Department of Physics and Astronomy, Johns Hopkins University,
Baltimore, MD 21218, USA}
\affil{$^2$D\'epartement d'Astronomie, Universit\'e de Gen\`eve, 51 Ch. des
Maillettes, 1290 Sauverny, Switzerland}
\affil{$^3$European Space Agency, European Space Astronomy Centre, P.O. Box 78,
Villanueva de la Ca\~nada, 28691 Madrid, Spain}
\email{ria@jhu.edu}

\altaffiltext{4}{Swiss National Science Foundation Fellow}

\begin{abstract}
Classical Cepheid variable stars are crucial calibrators of the cosmic distance
scale thanks to a relation between their pulsation periods and luminosities.
Their archetype, \object{$\delta$\,Cephei}, is an important calibrator for
this relation.
In this paper, we show that \DC\ is a spectroscopic binary based on
newly-obtained high-precision radial velocities. We
combine these new data with literature data to determine the orbit, which has
period $2201$\,days, semi-amplitude $1.5$\,\kms , and high eccentricity
($e=0.647$). We re-analyze {\it Hipparcos} intermediate astrometric data to measure
\DC 's parallax ($\varpi=4.09 \pm 0.16$\,mas) and find tentative evidence for an
orbital signature, although we cannot claim detection. We estimate that
{\it Gaia} will fully determine the astrometric orbit. Using the available
information from spectroscopy, velocimetry, astrometry,  
and Geneva stellar evolution models ($M_{\delta \rm{Cep}} \sim
5.0 - 5.25$\,\Msol), we constrain the companion mass to within $0.2 < M_2 <
1.2$\,\Msol . We discuss the potential of ongoing and previous interactions between the companion and \DC\ near pericenter passage, informing reported observations of circumstellar material and bow-shock. The orbit may
have undergone significant changes due to a Kozai-Lidov mechanism driven by the
outer (visual and astrometric) companion HD\,213307. 
Our discovery of \DC 's nature as a spectroscopic
binary exposes a hidden companion and reveals a rich and dynamical history of
the archetype of classical Cepheid variables. 
\end{abstract}


\section{Introduction}
Classical Cepheid variable stars (from hereon: Cepheids) are precise
Galactic and extragalactic distance tracers and thus of crucial importance for
cosmology.
The prototype of this class of stars, \DC\ (\object{HD 213306}, \object{HIP
110991}), has been extensively\footnote{SIMBAD (\url{http://simbad.u-strasbg.fr}) lists more than 600 articles
related to \DC .} studied ever since the discovery of its variability 230
years ago by \cite{1786RSPT...76...48G}. Until Baade's proposed method
\citep{1926AN....228..359B} to test Shapley's pulsation hypothesis
\citep{1914ApJ....40..448S} bore fruit in the early to mid 20th century, Cepheids were thought to be binary stars on eccentric orbits
\citep[see e.g.][]{1997VA.....41...95G}, inspiring a great deal of research, not
least that by Christian Doppler.

The first scientists to measure the variability of \DC 's spectral lines and to
determine their velocities were
\citet{1894AN....136..281B,1895ApJ.....1..160B} and \citet{1913LicOB...7..153M},
before radial velocities (RVs) became available for many Cepheids thanks
to the observations by \citet{1937ApJ....86..363J}. Two decades later,
\citet{1958ApJ...127..573S} conducted a detailed analysis of \DC 's RV curve and
concluded that no evidence of long-period changes in the mean velocity could be
seen, a result that was confirmed much later by \cite{1990ApJ...351..606K}. 

Nowadays, Cepheids are known to be pulsating variable stars, although many
Cepheids are also known to be binaries\footnote{cf.
the Cepheid binary database by \citet{2003IBVS.5394....1S} at
\url{http://www.konkoly.hu/CEP/nagytab3.html}}. The binary fraction of
Cepheids is being studied intensively with most recent estimates of the total
binary fraction ranging around $60\%$, see e.g.
\citet{2013AJ....146...93R} and \citet{2013MNRAS.434..870S}. However, Cepheids
cannot reside in very close-in binary systems
\citep[e.g.][]{2014arXiv1412.3468N} due to their nature as evolved (super-)giant
stars, which results in observed minimum orbital periods of approximately one
year. For long periods ($> 10$\,years), practical constraints create significant
observational bias against companion detection.

To our knowledge, \DC\ has not been shown to be
a spectroscopic binary prior to this work.
\DC\ is, however, a known \emph{visual} binary \citep{1966AJ.....71..119F},
whose companion HD\,$213307$ ($=$ HIP\,$110988$) is itself is an astrometric
binary and thought to be physically associated \citep[see][and references
therein]{2002AJ....124.1695B}.
HD\,$213307$ was proposed to also be a spectroscopic binary (Herbig, priv. comm.
mentioned in \citealt{1966AJ.....71..119F}). Finally, \DC\ is usually considered
to be a member of the loose association Cepheus\,OB6
\citep{1999AJ....117..354D,2007MNRAS.379..723V,2012ApJ...747..145M}.

Precise trigonometric parallax of \DC\ has been measured by
\citet{1997ESASP1200.....P}, \citet{2007MNRAS.379..723V}, and
\citet{2007MNRAS.379..723V} using observations made by the {\it Hipparcos}
space mission and by \citet{2002AJ....124.1695B} using measurements 
obtained with FGS\,3 on board the Hubble space telescope ({\it HST}).
Notable previous distance estimates include those reported by
\citet{1989MNRAS.237..947F},
\citet{1993ApJ...418..135G}, \citet{1993PASP..105.1101G},
\citet{1997A&A...317..789M}.

\citet{2005A&A...438L...9M} employed infrared long-baseline interferometry
to study the Baade-Wesselink projection factor of \DC\ and were later 
\citep{2006A&A...453..155M} able to show the presence of an extended
circumstellar envelope. The presence of this circumstellar environment was
confirmed independently (using different methodologies) by
\citet{2010ApJ...725.2392M} and \citet{2012ApJ...744...53M}.

In this paper, we present the discovery of the spectroscopic binary nature of
\DC . This discovery is demonstrated using new
observations that are presented in Sect.\,\ref{sec:newobs}. Using these new
data we reveal the presence of a hidden companion in
Sect.\,\ref{sec:hermes}. After combining our new RVs with literature data in
Sect.\,\ref{sec:literature}, we determine the orbital solution in
Sect.\,\ref{sec:orbit}. We re-analyze the {\it Hipparcos} intermediate
astrometric data from the \citet{2007ASSL..350.....V} reduction in Sects.\,\ref{sec:HIPpar}
and \ref{sec:HIPorbit} to measure parallax and to investigate whether the
available astrometric measurements are sensitive to the motion due to binarity. In
Sect.\,\ref{sec:Gaia}, we investigate {\it Gaia}'s expected sensitivity to
the binary motion. We discuss how our discovery helps to better interpret
other observations and begins to draw a complex picture of \DC 's rich and
dynamical history in Sect.\,\ref{sec:puzzle} before concluding in 
Sect.\,\ref{sec:conclusions}.

\section{New \hermes\ Observations}\label{sec:newobs}
Line-of-sight (radial) velocities were measured from $136$ observations taken
between September 2011 and September 2014 using the fiber-fed high-resolution (R
$\sim$ 85\,000) spectrograph \hermes\ \citep{2011A&A...526A..69R} at the Flemish
1.2m Mercator telescope located at Roque de los Muchachos Observatory, La Palma,
Canary Islands. We utilize the
high-resolution fiber (HRF) mode for all observations, since this is the most
commonly used, i.e., best-understood, observing mode available for \hermes . The
HRF mode offers optimal efficiency and the highest available spectral
resolution. 

The reduction pipeline available for \hermes\ performs 
pre- and overscan bias correction, flatfielding using Halogen lamps, and
background modelization, as well as cosmic ray removal. ThAr lamps are used
for the wavelength calibration. RVs are determined via the cross-correlation
technique \citep{1996A&AS..119..373B,2002A&A...388..632P} using a
numerical mask designed for solar-like stars (optimized for spectral type G2).

These observations were started as part of a 
search for Cepheids belonging to open clusters \citep{2013MNRAS.434.2238A} with the 
goal of quantifying the precision limit for a
classical Cepheid using \hermes\ RVs and of having high-quality spectra
available for further study. The average signal-to-noise ratio of the spectra is
higher than 200 near $6000$\,\AA, with some spectra reaching up to $400$.

Table\,\ref{tab:HermesRVs} shows a sample of our \hermes\ RV measurements that
we are making publicly available at the
CDS\footnote{\url{http://cds.u-strasbg.fr/}} together with our standard
star observations.

\begin{table}
\centering
\begin{tabular}{llrr}
\hline 
BJD - $2\,400\,000$ & Phase & $v_r$ & $\sigma ( v_r )$ \\
 &  & [ \kms ] & [ \kms ] \\
\hline
$55816.469627$ & $0.82074$ & $0.802$ & $0.018$ \\
$55816.470920$ & $0.82098$ & $0.813$ & $0.018$ \\
$55816.472221$ & $0.82122$ & $0.824$ & $0.018$ \\
$55817.510667$ & $0.01474$ & $-21.125$ & $0.018$ \\
$55817.511961$ & $0.01498$ & $-21.202$ & $0.018$ \\
$55818.494094$ & $0.19800$ & $-31.354$ & $0.018$ \\
$55818.495384$ & $0.19824$ & $-31.362$ & $0.018$ \\
$55818.496675$ & $0.19848$ & $-31.342$ & $0.018$ \\
$55819.497269$ & $0.38494$ & $-21.989$ & $0.018$ \\
$55819.498564$ & $0.38518$ & $-21.984$ & $0.018$ \\
\multicolumn{4}{c}{full table available at {\tt CDS}} \\
\hline
\end{tabular}
\caption{Sample of the new \hermes\ RV data that are published in their entirety
at the {\tt CDS}, and shown here for guidance regarding their form and content.
Radial velocities have been shifted to the {\it CORAVEL-ELODIE} zero-point.
Pulsation phase is defined here as 0 when $v_r = v_\gamma$ at the steep part of
the RV curve. Uncertainties are fixed at $18$\,\ms\ to account for
uncertainty in the wavelength calibration and zero-point calibration, see text.}
\label{tab:HermesRVs}
\end{table}

\subsection{Zero-point, Stability, and Precision of \hermes\
RVs}\label{sec:hermeszp}

To determine the RV zero-point of \hermes , we observed eight RV standard stars
listed in \citet{1999ASPC..185..383U,1999ASPC..185..367U}\footnote{see
\url{http://obswww.unige.ch/$\sim$udry/std/std.html}}, see
Tab.\,\ref{tab:RVstds}. These standard stars were chosen to cover the range of
spectral types that \DC\ exhibits during its pulsation cycles. We thus determine
a mean systematic offset of $55$\,\ms\ with respect to the {\it ELODIE} and {\it
CORAVEL} zero-point. We estimate this zero-point offset to be accurate to
approximately $10$\,\ms\ and note that additional scatter in the difference
between \hermes\ and literature RVs can exist for various reasons of
astrophysical origin, including binarity and planetary companions.
We therefore increase our error margin by adding $10$\,\ms\ in quadrature to the
\hermes\ RV uncertainties when correcting for zero-point offsets.

\begin{table}
\centering
\begin{tabular}{llrrrrr}
\hline
HD    & Sp.Type & $\langle v_r \rangle$ & $v_{r\rm{,ref}}$ & $\Delta
v_{r{\rm{,ref}}}$ & RMS$_{\rm{nc}}$ & RMS$_{\rm{corr}}$ \\
      & & [\kms ]  & [\kms ] &  [\kms ]  &  [\ms ]  & [\ms ] \\ \hline   
10780 & K0V & 2.771    & 2.70  &  0.071 &   38.7 & 17.1    \\ 
32923 & G4V & 20.627   & 20.50  & 0.127 &    70.1 & 14.8   \\
42807 & G2V & 6.068    & 6.00 &  0.068 &  115.2 & 29.7 \\
82106 & K3V & 29.784   & 29.75 & 0.034  &   45.8 & 20.8  \\
144579 & G8V &  -59.452 & -59.45 & -0.002   & 42.7 & 20.0   \\
168009 & G1V &  -64.581 & -64.65 & 0.069  &  36.4 & 20.0   \\
197076 & G5V &  -35.413 & -35.40 & -0.013  &  78.3 & 10.9   \\
221354 & K0V &  -25.111 & -25.20 & 0.089  &  53.8 & 15.5  \\
\hline
\end{tabular}
\caption{List of RV standard stars with spectral types from
SIMBAD, \hermes\ RVs of standard stars from new observations,
\citet{1999ASPC..185..367U}, offsets between new and reference RVs, and RMS of new observations without (subscript nc) and with pressure corrections applied
(subscript corr) following \citet[Sec.\,2.1.5]{2013PhDT.......363A}.}
\label{tab:RVstds}
\end{table}

We investigate the long-term stability and precision of \hermes\ RVs using
the RV standard star observations. Thanks to the very high signal-to-noise ratio
of our spectra, photon-noise
\citep{2001A&A...374..733B} contributes only marginally to the uncertainty of
our \hermes\ RVs. Instead, the precision of our RVs is dominated by the
intra-night stability of the wavelength calibration and the long-term stability
of the instrument.

To improve RV precision and stability, we employ a method developed in
\citet[Sec.\,2.1.5]{2013PhDT.......363A} to correct for the RV drift
induced by atmospheric pressure variations that occur during the night. Pressure variations are the
leading cause for drifts in \hermes\ RVs, since the temperature of the
instrument is stabilized to within $0.01^\circ$\,K \citep{2011A&A...526A..69R}.
The RV drift correction due to pressure variations is based on changes in the
refractive index of air and has been shown to be precise to approximately
$10$\,\ms\ in the case of the {\it Coralie} spectrograph\footnote{{\it Coralie}
is mounted to the Swiss 1.2m Euler telescope located at La Silla Observatory,
Chile} for which RV drift corrections are measured using interlaced
simultaneous ThAr exposures. For \hermes , our method improves RV precision by a factor of 
approximately $2.5$, as measured by the decrease in RMS for all standard stars.

\begin{figure}
\centering
\includegraphics[scale=0.75]{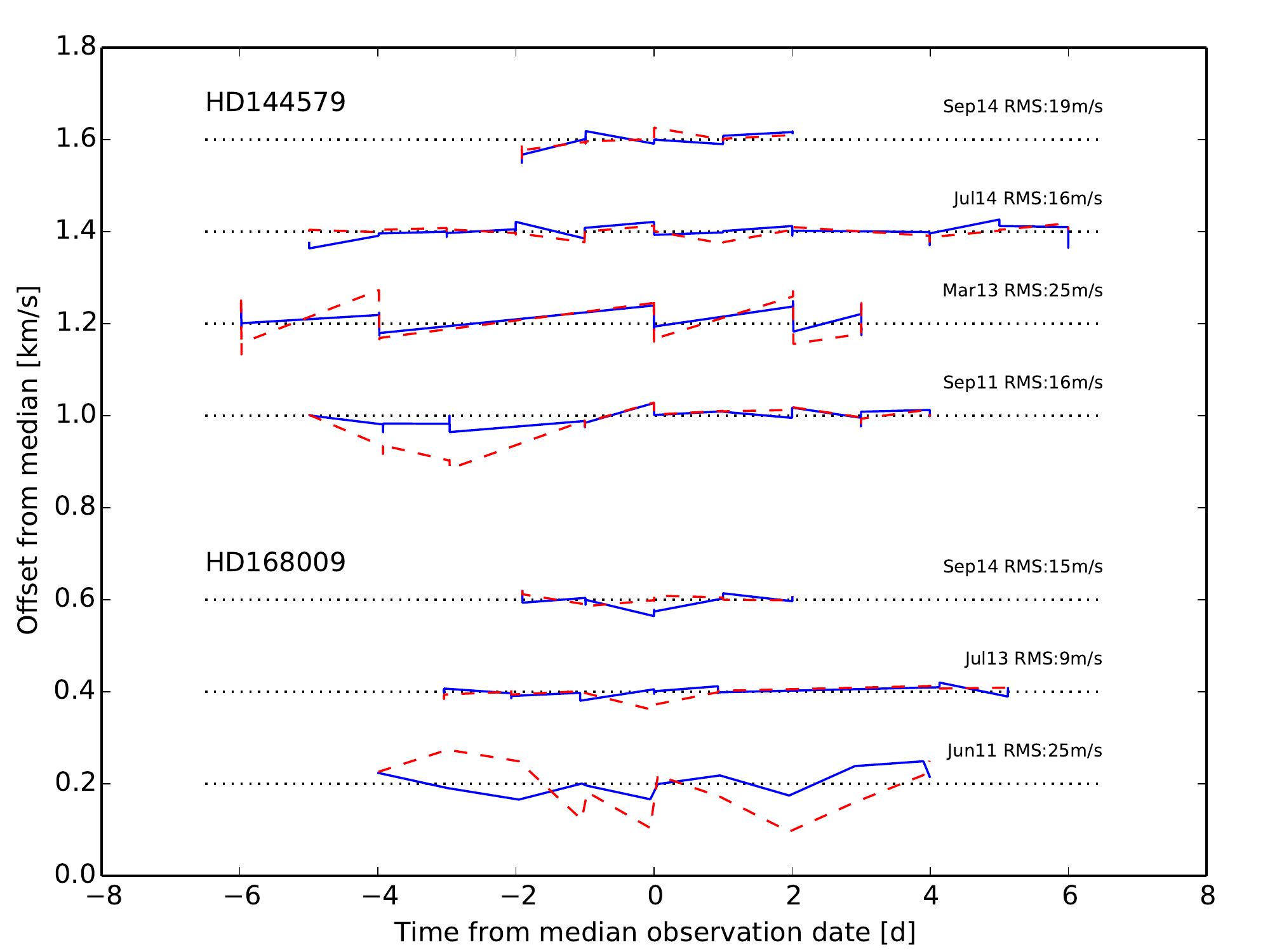}
\caption{Hermes RV standard stars HD\,144579 and HD\,168009. The measurements
are separated into individual observing runs and plotted as a function of time
centered around the median observation date of a given star in a given
observing run.
We plot the measurements as residuals around the median of all measurements
(i.e., including data from all observing runs) as listed in
Tab.\,\ref{tab:RVstds}.
We offset measurements from different observing runs and different stars (black
dotted line) from one another for clarity.
Pressure-corrected (see text) measurements are shown as solid blue connected
dots, the uncorrected measurements are shown as a red dashed line. The labels on
the right indicate observing run and RMS of the pressure-corrected RVs.}
\label{fig:HermesRVstds1}
\end{figure}

Figure\,\ref{fig:HermesRVstds1} shows both the uncorrected (red dashed lines)
and corrected RVs (blue solid line) for the standards stars HD\,144579 and
HD\,168009 as a function of time for the different observing runs during which
they were observed.
The highest precision of $9$\ms\ is achieved for HD\,168009 during a ten-night
observing run in July 2013. We find no evidence for long-term variations over
the 3 year duration of the observations.
Based on all standard star measurements, we estimate the long-term precision of
pressure-corrected \hermes\ RVs to be approximately $15$\ms . Accounting for
the additional uncertainty due to zero-point offsets (see above), we adopt $18$\,\ms\ as our
error budget for the investigations based purely on \hermes\
RVs\footnote{However, the uncertainty adopted to determine \DC 's orbit
is $47$\,\ms , cf. Sect.\,\ref{sec:hermes} and Fig.\,\ref{fig:hermesonly}.}.
Note that this adopted uncertainty of $18$\,\ms\ is more than a
factor $2000$ smaller than the pulsation-induced peak-to-peak RV amplitude of
$38.6$\,\kms.

\section{Results from Spectroscopy}\label{sec:spectroscopy}
In this section we describe our analysis of \DC 's RV curve, starting with the
initial discovery of \DC 's nature as a spectroscopic binary
(Sect.\,\ref{sec:hermes}). After subtracting the RV drift due to orbital motion
observed in \hermes\ data, we establish a high-precision reference model for the
pulsations alone, which we subtract from a combined data set of 
RVs from \hermes\ and several literature sources (cf.
Sect.\,\ref{sec:literature}).
Using this combined dataset, we determine the orbit for the binary system
in Sect.\,\ref{sec:orbit}.

We also searched the individual high-quality \hermes\ spectra as
well as the cross-correlation profiles for a signature identifying the
companion, finding none. We note that previous analyses of
IUE \citep{1993AJ....106..726E} and HST/COS \citep{2014ApJ...794...80E} spectra
did not report evidence of a companion's signature, effectively ruling out
early-type companions. Given the high signal-to-noise ratio (S/N) of our spectra
and the even better S/N of the cross-correlation profiles, we estimate that the 
companion must be at least a factor  $100$ fainter than \DC , at least in
optical bandpasses.
Based on Geneva stellar evolution models \citep{2013A&A...553A..24G} and
assuming an approximately $5$\,\Msol\ Cepheid, this implies a 
companion mass below approximately $1.75$\,\Msol . This
is consistent with the upper limit on companion spectral type (A3, i.e.,
$M_2 < 2$\Msol) set by non-detection in IUE spectra \citep{1992ApJ...384..220E}.

\subsection{Discovery of \DC\,B using HERMES RVs}
\label{sec:hermes}
\begin{figure}
\centering
\includegraphics[scale=0.8]{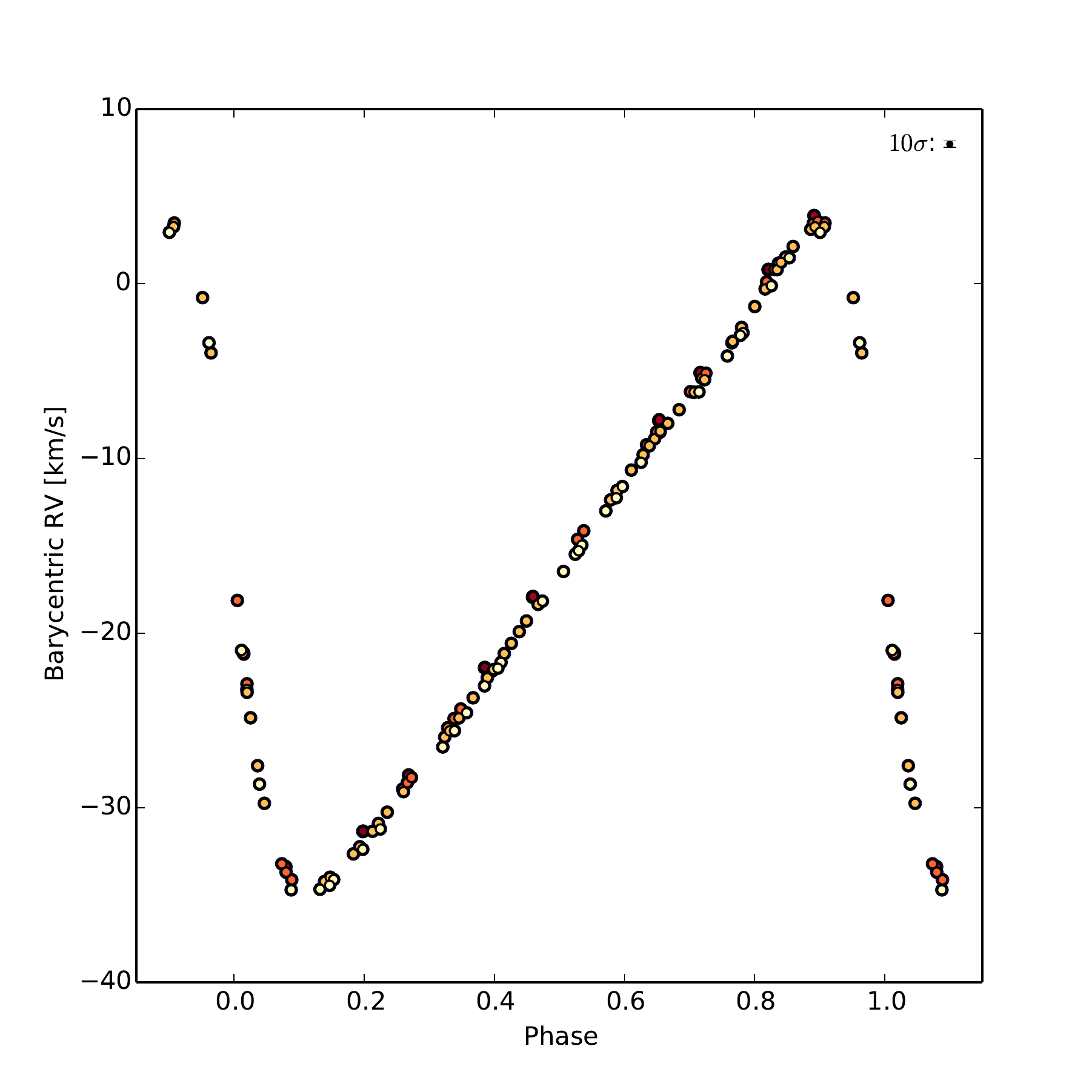}
\caption{Phase-folded HERMES radial velocity curve with
$P_{\rm{puls}}=5.366274$\,d. Observation date is traced by a color
scale and increases from red to yellow. The
ten-fold mean uncertainty of the measurements is shown in the top right corner.
The RV offset at constant phase is due to the spectroscopic binary nature of \DC .} 
\label{fig:hermesRVphased}
\includegraphics[scale=0.8]{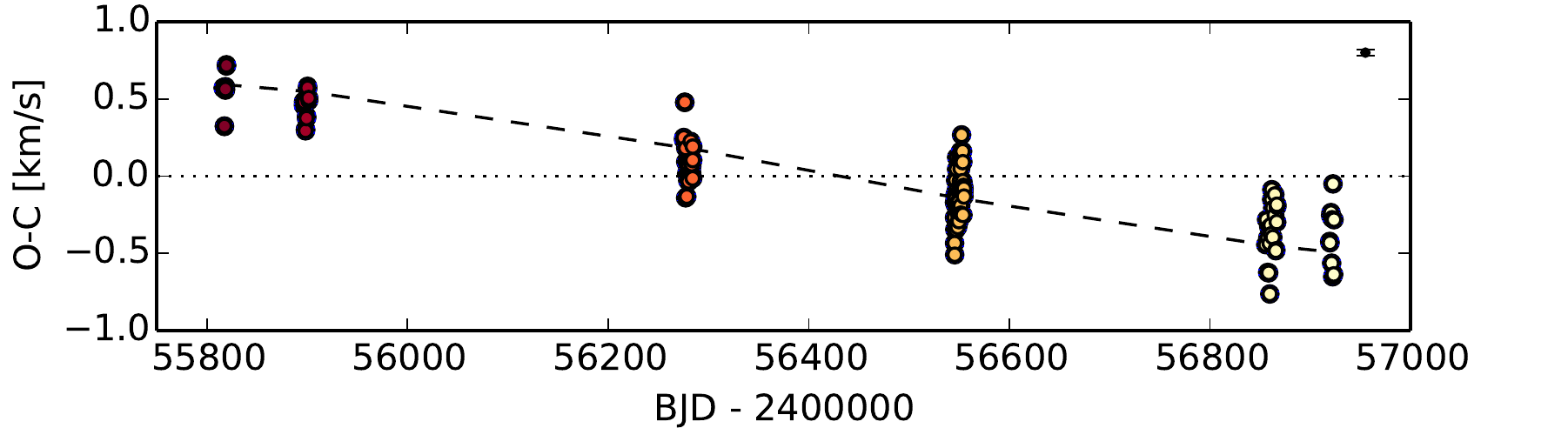}
\caption{Residuals for \hermes\ RVs minus the pulsation model as a function of 
observation date. The trend (dashed line) reveals the presence of \DC 's 
spectroscopic companion.}
\label{fig:hermesRVphasedresiduals}
\end{figure}

Figure\,\ref{fig:hermesRVphased} shows the \hermes\ RVs obtained for \DC ,
phase-folded with the best-fitting pulsation period of $P_{\rm{puls}} =
5.366274$\,d.
The ten-fold RV uncertainty is shown in the upper right corner.
The observation date is traced by symbol color going from red (first
observations in September 2011) to yellow (September 2014).
The color coding reveals the average radial velocity to be time-dependent,
thus revealing orbital motion caused by a hidden companion.

We model the RV variations due to pulsation using the technique described in
\citet{2013MNRAS.434.2238A}. In a nutshell, we fit a Fourier series to model the
phase-folded RV curve and increase the number of harmonics until an F-test
indicates spurious fit improvement. This approach yields fit residuals that
clearly exhibit a strong temporal correlation with an RMS of
$337$\,\ms, see Fig.\,\ref{fig:hermesRVphasedresiduals}. 

By running the fitting algorithm while assuming a model composed of the
sum of a Fourier series and linear, quadratic, and cubic trends, we find that
the orbital drift seen in \hermes\ RVs is best described by a cubic polynomial
and that the best-fit pulsation model has 14 harmonics. We retain this model as
our pulsation reference model for the following steps.

\begin{figure}
\centering
\includegraphics{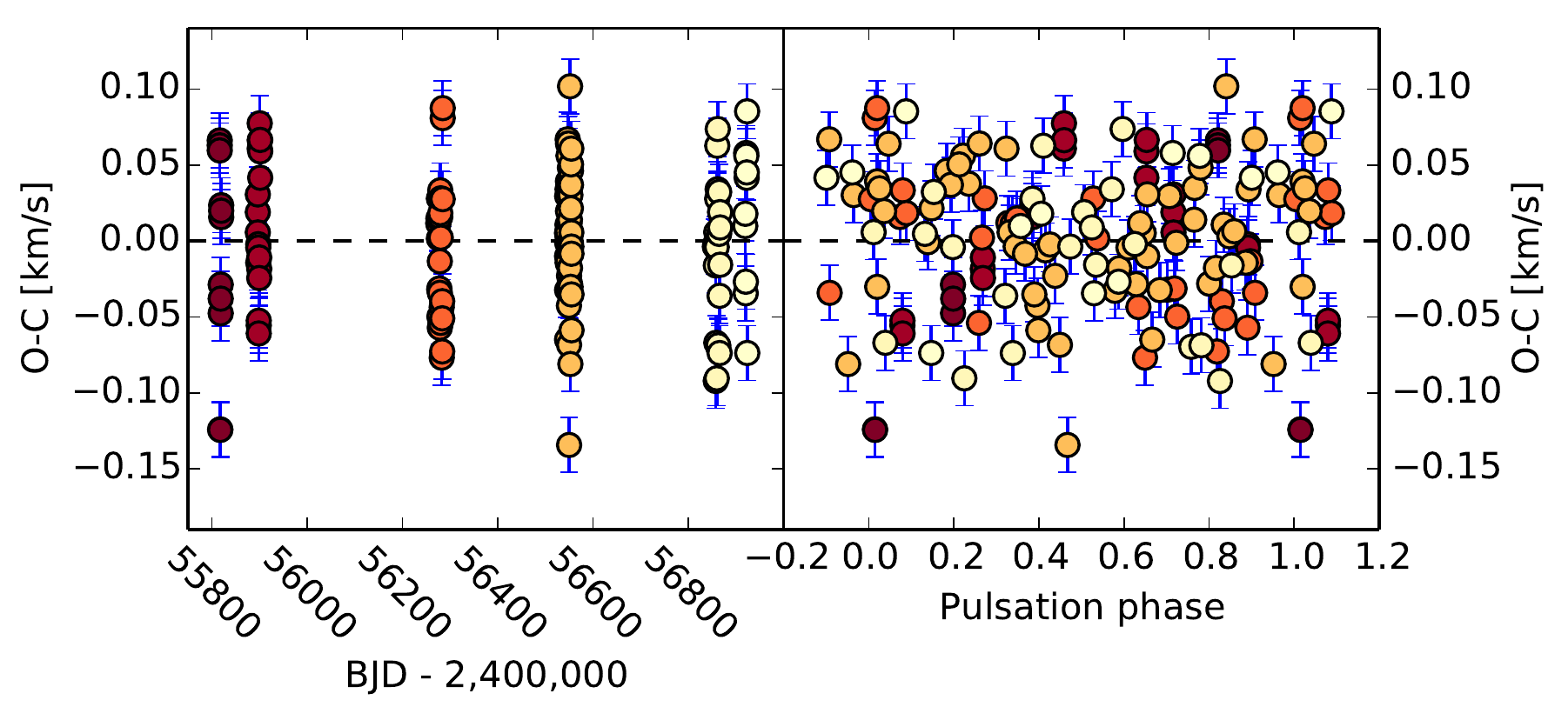}
\caption{\hermes\ RV residuals (RMS$= 47$\,\ms) after accounting for both
pulsation and the cubic drift (see text) due to binarity. The
left panel shows residuals as a function of the observation date, the right
panel as a function of pulsation phase, with $P_{\rm{puls}}=5.366274$\,d. We
trace observation date by color scaling the data points (red is
oldest, yellow is newest) to expose additional signal that is not
correctly modeled by pulsation and binarity and is likely related to random
fluctuations in pulsation period (period-jitter).} 
\label{fig:hermesonly}
\end{figure}

Figure\,\ref{fig:hermesonly} shows the residuals from \hermes\ RVs after
subtracting the model that accounts for pulsations as well as the cubic
drift due to orbital motion.
The residuals are flat over the observational baseline. However, the
phase-folded residuals do exhibit some structure, which is exposed by
applying a color scale to trace the observation date. The RMS of our
\hermes\ residuals is $47$\,\ms , i.e., higher than the $18$\,\ms\ estimated from
RV standard stars in Sec.\,\ref{sec:hermeszp}.

This higher-than-expected residual may be explained by small stochastic
variations in the pulsation period seen in photometry of other Cepheids
obtained with the {\it Kepler} and {\it MOST} satellites \citep[so-called
period-jitter, see][]{2012MNRAS.425.1312D,2015MNRAS.446.4008E}, which has been explained as
being due to surface convection and granulation \citep{2014A&A...563L...4N}. We
note also that \citet{2014A&A...566L..10A} recently
discovered RV curve modulation in Cepheids, albeit at much lower amplitudes.
Since these effects (period-jitter or modulation) limit our ability to
precisely reproduce the pulsation curve, we adopt the RMS value of $47$\,\ms\ as
our \hermes\ RV uncertainty for \DC\ for the remainder of our analysis.

\subsection{Combination with literature data}\label{sec:literature}
The cubic drift seen in the residuals in
Fig.\,\ref{fig:hermesRVphasedresiduals} indicates orbital motion at a timescale
longer than the observational baseline achieved by our \hermes\ observations. We
therefore searched the literature for data suitable for determining the
orbit of \DC .

\DC\ is one of the most-studied variable stars and several authors have
previously published RV data for it, including \citet{1958ApJ...127..573S},
\citet{1987ApJS...65..307B},
\citet{1989ApJS...69..951W},
\citet{1993ApJ...415..323B}, \citet{1994A&AS..108...25B},
\citet{1996AstL...22..175G}, \citet{1998MNRAS.297..825K}, \citet{2004A&A...415..531S}, \citet{2005ApJS..156..227B}.
Historically, \DC 's RV curve has been thought to be well-understood, which may
in part explain why its spectroscopic binary nature has gone unnoticed for so
long. The main reason, however, is that \DC 's orbital signature has a small RV
amplitude, high eccentricity, and long orbital period, and thus requires
high-precision velocimetry over an observational baseline spanning at least two
years. For comparison, the various RV datasets available in the literature have
typical observational baselines on the order of one year and do not have
sufficient precision to detect binarity during this timeframe.

Zero-point offsets must be corrected for when combining RV data from
different instruments and the literature. \DC\ represents a particularly
difficult case, because zero-point differences can be on the
same order of magnitude as the difference in $v_\gamma$ due to orbital motion
between datasets. Therefore, a well-determined common zero-point is the
key to determining the orbit of \DC\ accurately. To this end, we
adopted the {\it CORAVEL-ELODIE} RV zero-point (cf.
\citealt{1999ASPC..185..383U,1999ASPC..185..367U} and
Tab.\,\ref{tab:RVstds}).

Of the available literature sources, we excluded the following from our
analysis due to insufficient precision ($\sigma_{v_r} > 1$\kms):
\citet{1958ApJ...127..573S}, \citet{1987ApJS...65..307B},
\citet{1989ApJS...69..951W}. We further discarded the measurements by
\citet{1993ApJ...415..323B} and \citet{2004A&A...415..531S}, since no
information is available to determine the zero-point differences with {\it
CORAVEL}. Finally, we excluded the single measurement published by
\citet{1996AstL...22..175G}.

Conversely, the measurements published by \citet{1994A&AS..108...25B} and
\citet{2005ApJS..156..227B}, as well as publicly available RV data from the {\it
ELODIE} archive\footnote{\url{http://atlas.obs-hp.fr/elodie/}}
\citep{2004PASP..116..693M} are already on the common {\it CORAVEL-ELODIE} RV
zero-point. We found that an offset of $-0.35$\,\kms\ to the measurements by
\citet{1998MNRAS.297..825K} is appropriate to improve agreement with the
(contemporaneous) data from the {\it ELODIE} archive and
\citet{2005ApJS..156..227B}.
This offset is similar to the precision stated by \citet[$\sim
0.3$\,\kms]{1998MNRAS.297..825K}. The determination of the
zero-point offset between \hermes\ and {\it CORAVEL-ELODIE} is discussed in
Sect.\,\ref{sec:hermes} above. Having thus calibrated the zero-point
differences based on measurements of standard stars, we can use the combined
data set to determine the orbit with confidence.

Classical Cepheids are known to exhibit changing periods due to their secular
evolution. However, Cepheids can also exhibit erratic changes of unknown
origin in their pulsation periods \citep[see e.g.][]{2000NewA....4..625B}. When
combining RV data from the literature, we noticed that variable periods
have to be accounted for.
We attempted to use the ephemerides and rate of (pulsation) period change by
\citet{2000ASPC..203..244B} to obtain accurately phase-folded RV
curves, but we were unable to obtain a satisfactory result. 

Since obtaining good phase-folding is required in order to correctly subtract
the pulsation reference model, we phase-folded the combined data set in the
following way.
First, we separated the dataset into three parts with different pulsation
periods. The motivation for separating the evolution of the pulsation period in
this way is the observation that the period of \DC\ changes very
slowly (\citealt{1919Obs....42..338E}, \citealt{2000ASPC..203..244B}). We then
determined the best-fit pulsation periods for each of these three epochs by
minimizing the scatter in the residuals after fitting for the pulsation alone.
We thus adopt the following pulsation periods:
$5.3657 \pm 0.0013$\,d for data by \citet[mean epoch JD
$2\,444\,467.12$]{1994A&AS..108...25B}; $5.36615 \pm 0.0005$\,d for data from
{\it ELODIE}, \citet{1998MNRAS.297..825K}, and \citet[mean epoch JD
$2\,450\,398.61$]{2005ApJS..156..227B}; $5.366274 \pm 0.00006$\,d for \hermes\
RVs (mean epoch JD $2\,456\,430.38$). While the center values of this sequence
would imply a slowly increasing pulsation period ($dP/dt \sim 0.5 - 1.1 \times
10^{-5}\,\rm{s\,yr^{-1}}$), these adopted periods agree to within their
uncertainties.
After applying our updated pulsation periods, we shifted all three epochs for
$\phi \equiv 0$ to occur at minimum radius, i.e., when the velocity is equal to
$v_\gamma$ on the steep part of the RV curve. This provides us with an
accurately phase-folded pulsation curve from which we subtract the pulsation
reference model to reveal the orbital motion of \DC .

\begin{figure}
\centering
\includegraphics{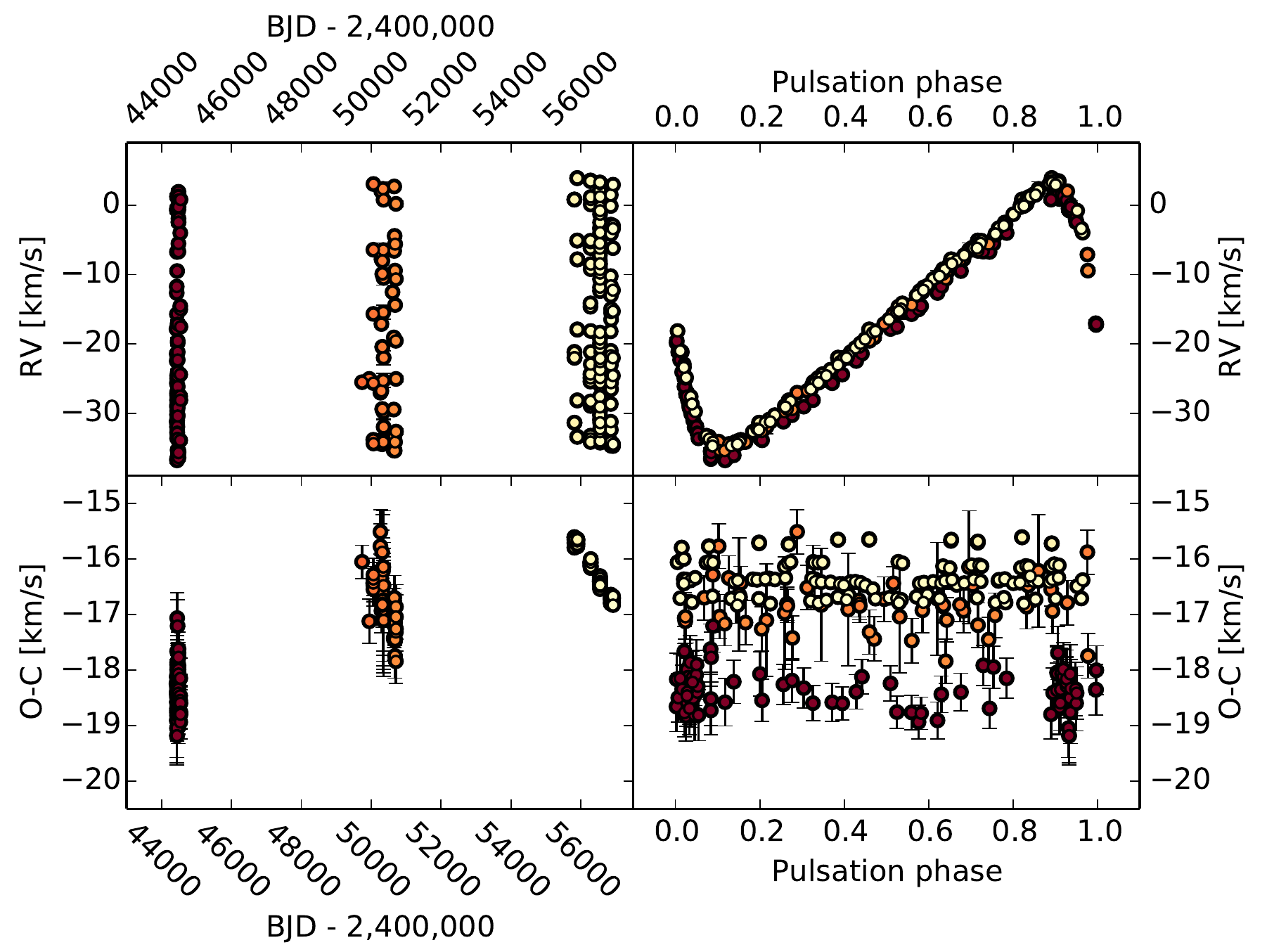}
\caption{The combined dataset (\hermes\ + literature RVs) modeled
only for pulsation-induced variability. The left panels show the measurements
(top) and residuals (bottom) against observation date, the right hand panels
show the same measurements against pulsation phase. Observation date is traced
by the applied color scale, with the newest measurements drawn white.}
\label{fig:combinedrv}
\end{figure}

Figure\,\ref{fig:combinedrv} shows the RV curve based on the combined dataset.
The residuals clearly demonstrate the presence of orbital motion.

\subsection{Orbit Determination using HERMES and Literature
RVs}\label{sec:orbit} 
\begin{figure}
\centering
\includegraphics[scale=0.8]{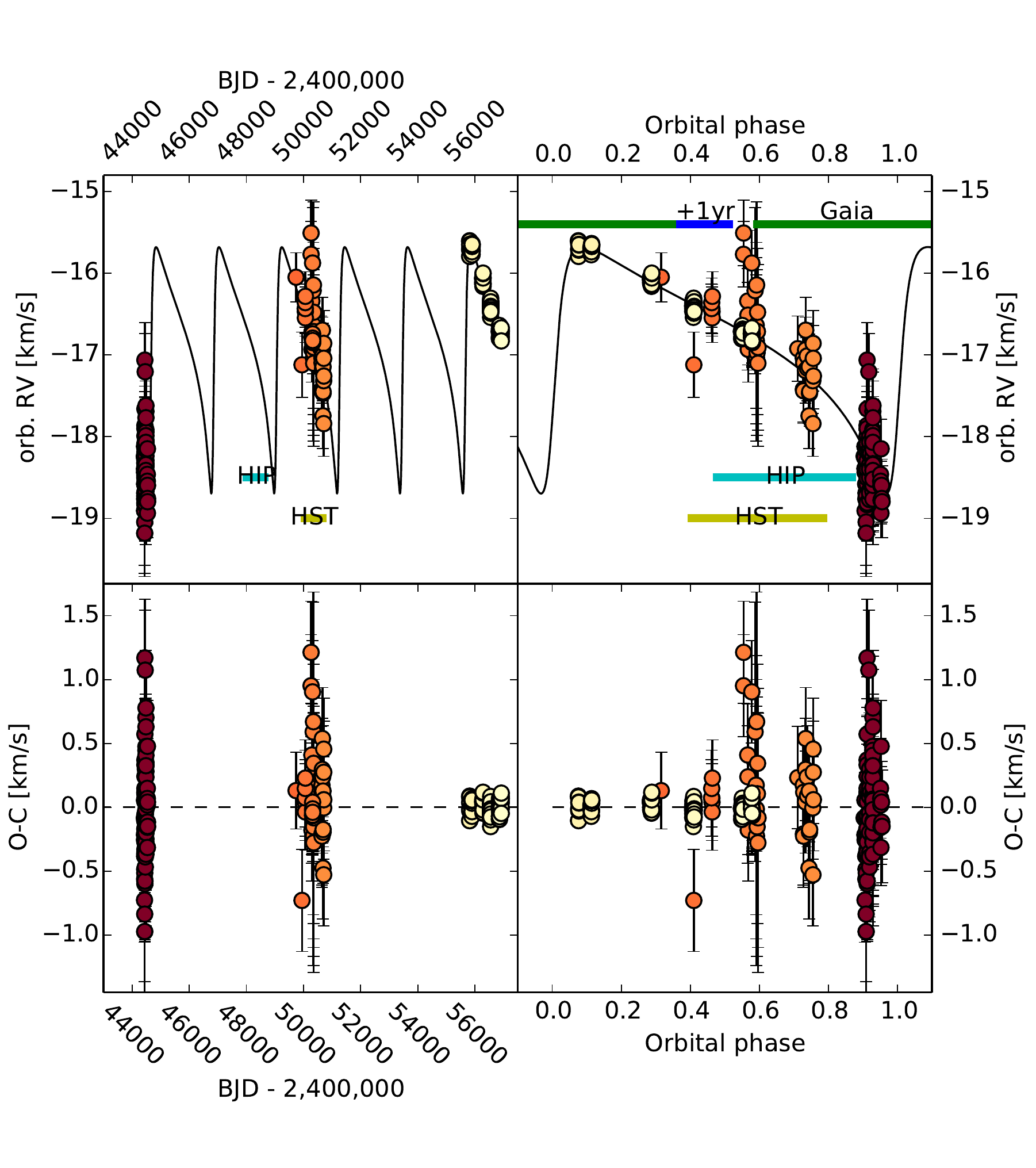}
\caption{Orbital solution for \DC\ based on the combined dataset from which we 
subtracted our pulsation reference model.
We trace observation date by color scaling data points from red
for the oldest to yellow for the newest measurements. Left panels show
measurements (top) and residuals (bottom) against observation date, right panels
show the same data against orbital phase, assuming $P_{\rm{orb}} = 2201.87$\,d, see
Tab.\,\ref{tab:orbitsolution}, and $\phi \equiv 0$ at pericenter passage.
We indicate the range of orbital phases at which {\it Hipparcos}, {\it HST}, and
{\it Gaia} have observed or will observe \DC .}
\label{fig:rvorbit}
\end{figure}

Figure\,\ref{fig:rvorbit} shows the orbital motion of \DC\
exposed by subtracting our pulsation reference model from the phase-folded
combined dataset. As can be seen in the top right panel, the orbit is
well-sampled except for the ascending part of the orbital RV curve.
Our observations did not sample this part of the RV curve, since 
it occurred shortly (a few months) before the start of our observations.

\begin{figure}
\centering
\includegraphics[scale=0.3]{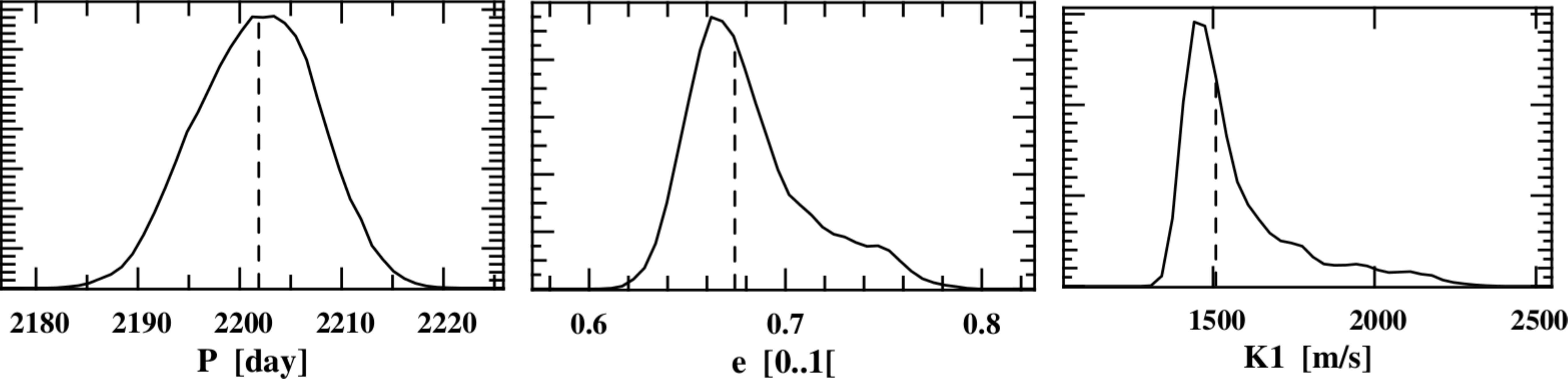}
\caption{Marginal probability density functions for orbital period,
eccentricity, and semiamplitude as determined by {\tt Yorbit}'s MCMC algorithm.
The dashed line indicates the median value adopted as numerical reference.}
\label{fig:rvorbitpdfs}
\end{figure}

Table\,\ref{tab:orbitsolution} provides the orbital solution for \DC\ determined
from the orbit-only combined RV curve using a standard Keplerian model.
We used the tool {\tt Yorbit} (S\'egransan, et al., in prep.) to first determine
an initial orbital estimate via a genetic algorithm and then characterize the parameter uncertainties using the marginal distributions of Markov Chain Monte Carlo simulations with
$500\,000$ iterations, cf. Fig.\,\ref{fig:rvorbitpdfs}. From the orbital
solution and assuming $M_{\delta \rm{Cep}} \sim 5.0 - 5.25$\,\Msol\ (cf.
Sec.\,\ref{sec:conclusions}) we determine the minimum mass of the companion to be $0.2
\pm 0.02$\,\Msol . While $a_{rel}$ and $a_1 \sin{i}$ are listed
assuming $M_{\delta \rm{Cep}} = 5.25$\,\Msol , the stated values remain within
the stated uncertainties if $M_{\delta \rm{Cep}} = 5.0$\,\Msol\ is adopted.

Comparing our results to
other known Cepheid orbits listed in the Cepheid binary database
by \citet{2003IBVS.5394....1S}, we find that the semi-amplitude $K$ of this
orbit is the second smallest among all known binary Cepheid orbits, albeit with
much larger eccentricity and somewhat longer period than W\,Sgr's orbit
\citep{2008A&A...488...25G}. This also helps to explain why \DC\ has not
previously been identified as a spectroscopic binary.

\begin{table}
\centering
\begin{tabular}{lrrrrrrrrr}
\hline
Parm & $v_\gamma$ & $T_0$ & $P_{\rm{orb}}$ & $K$ & $e$ & $\omega$ &
$a_{\rm{rel}}$ & $a_1 \sin{i}$ & $f_m$ \\
Unit & [\kms ] & [d] & [d] & [\kms ] & & [deg] & [au] & [$10^{-3}$\,au] &
[$10^{-3}$\,\Msol ] \\
\hline
Value & $-16.787$ & $55649.68$ & $2201.87$ & $1.509$ & $0.674$ & $246.77$ &
$5.82$ & $226.3$ & $0.784$ \\
$\sigma_+$ & $0.026$ & $24.68^\dagger$ & $5.73$ & $0.239$  &
$0.038$ & $2.37$ & $0.18$ & $21.9^\dagger$ & $0.249^\dagger$\\
$\sigma_-$ & $0.049$ & $19.86^\dagger$ & $6.31$ & $0.080$ & $0.021$ & $4.90$ &
$0.19$ & $7.9^\dagger$ & $0.083^\dagger$ \\
\hline
\end{tabular}
\caption{Orbital solution for \DC\ based on the combined \hermes\ and
literature radial velocities. $\sigma_+$ and $\sigma_-$ denote the upper and
lower standard errors derived from marginal distributions. Quantities with
superscript dagger have been computed using Gaussian error propagation. 
Other uncertainties were estimated by the MCMC analysis.}
\label{tab:orbitsolution}
\end{table}

\section{Searching for an Astrometric Orbital Signature}
Projecting the orbit derived in Sect.\,\ref{sec:orbit} to the distance
determined by \citet[$\varpi = 3.66 \pm 0.15$\,mas]{2007AJ....133.1810B} yields
an orbital relative semimajor axis of $a_{\rm{rel}} = 21.2$\,mas, with an
orbital barycentric semimajor axis of the Cepheid of $a_1 > 0.84$\,mas, a value
more than five times the uncertainty estimated by \citet{2007AJ....133.1810B},
and even seven times the uncertainty stated by \citet[{\it Hipparcos}, $3.71 \pm
0.12$\,mas]{2007MNRAS.379..723V}.
This prompts the question whether the observations taken by {\it Hipparcos} or
{\it HST} are sensitive to the orbital motion, cf.
Fig.\,\ref{fig:rvorbit} for information as to which ranges of orbital
phase were observed by these missions.
We therefore explore the sensitivity of {\it Hipparcos} and {\it Gaia}
astrometry to \DC 's hidden companion in this section.

\subsection{Hipparcos Intermediate Astrometric	 Data}
To test whether {\it Hipparcos} \citep{1997ESASP1200.....P} was sensitive to the
orbital motion of \DC , we analyze the {\it Hipparcos} intermediate astrometric
data (IAD) published on the DVD attached to the new reduction by
\citet{2007ASSL..350.....V}.

\subsubsection{Parallax from Intermediate Astrometry Data}\label{sec:HIPpar}
\begin{figure}
\centering
\includegraphics[scale=0.5]{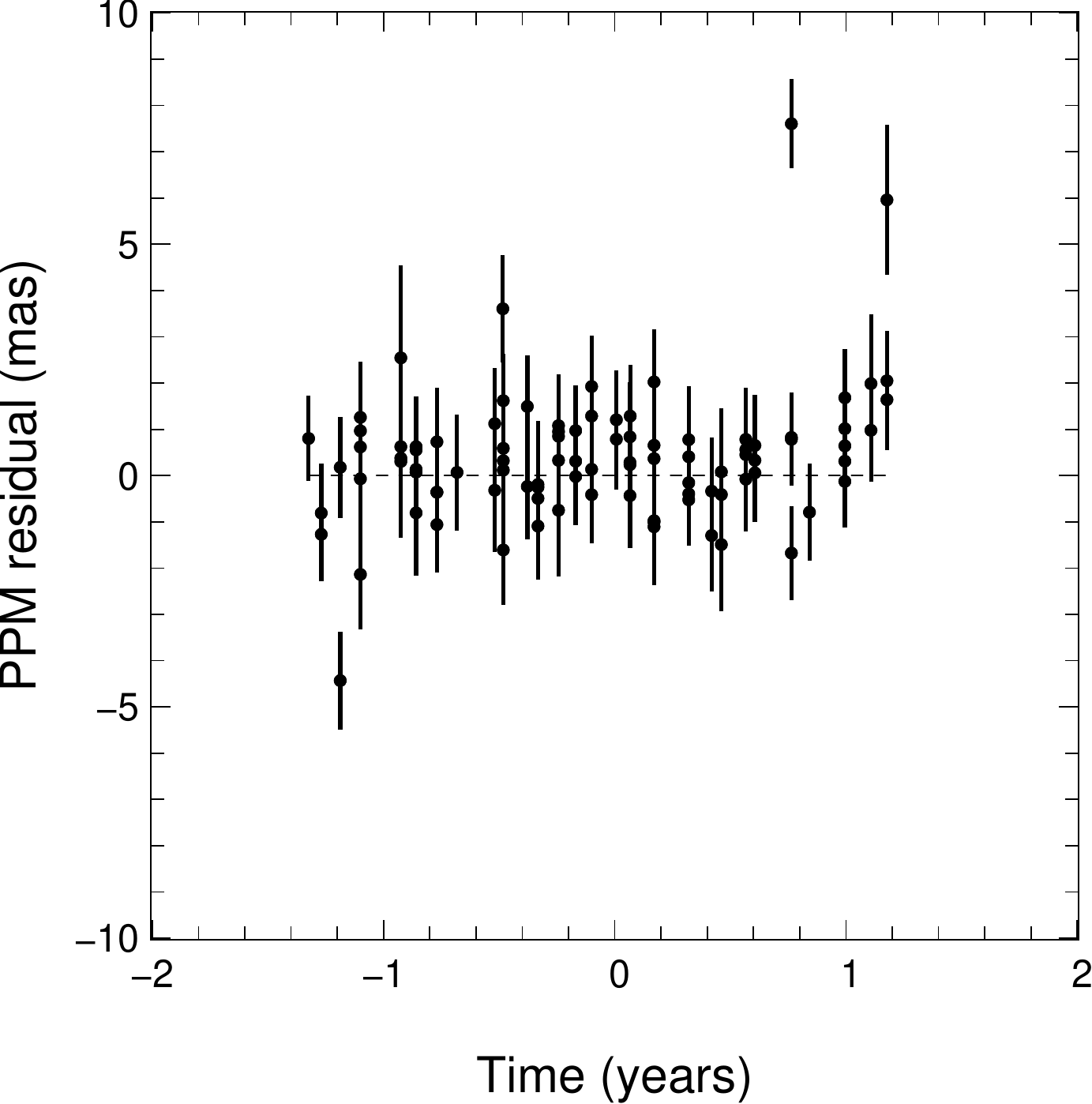}
\includegraphics[scale=0.5]{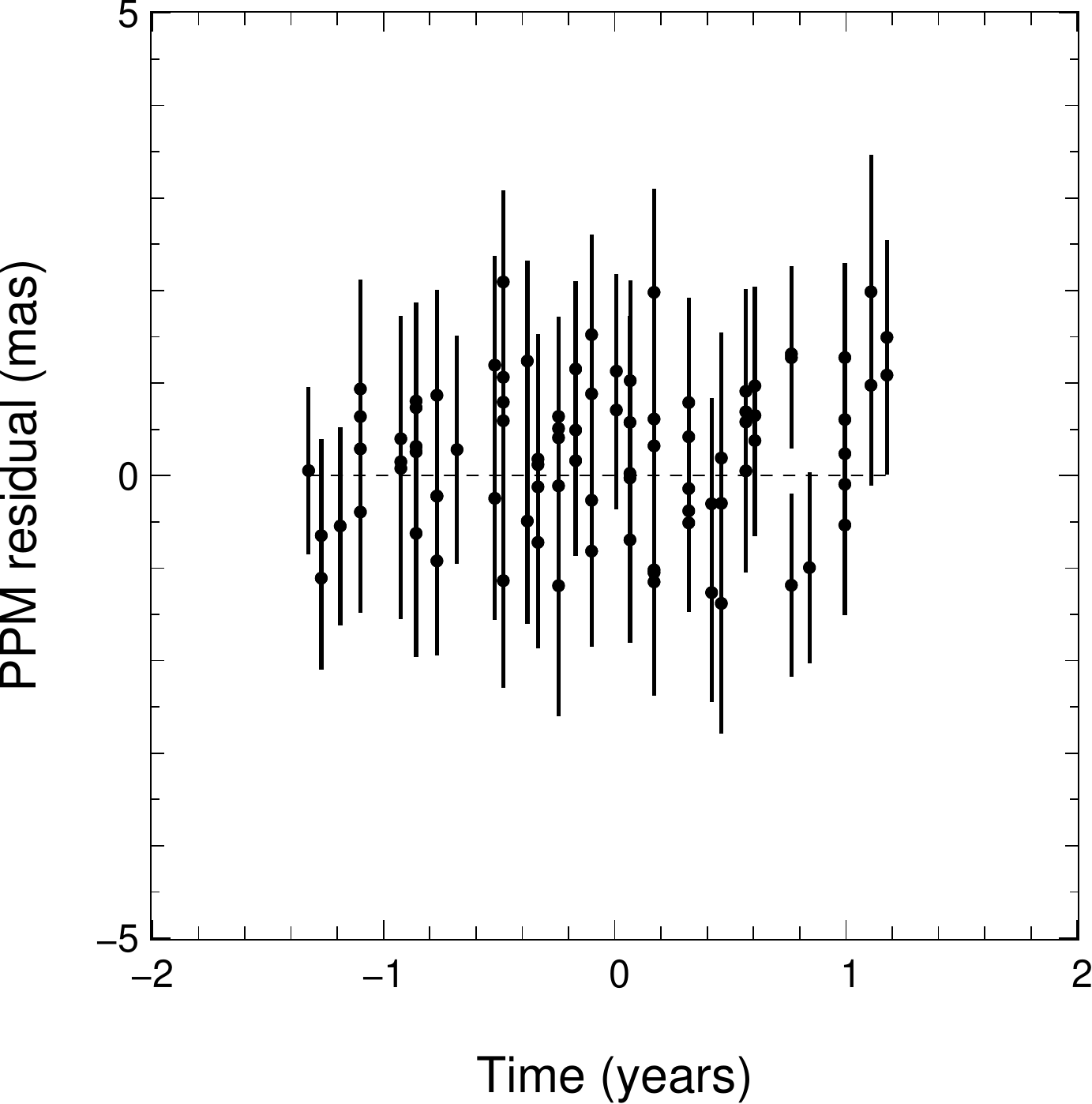}
\caption{{\it Hipparcos} fit residuals for all 95 data points (left panel) and after removing 6 outliers
(right panel). Time is relative to epoch J$1\,991.25$ as in the original
{\it Hipparcos} catalog \citep{1997ESASP1200.....P}.}
\label{fig:HIPresiduals}
\end{figure}

A 5-parameter fit to the $95$ measurements given on the DVD yields the residuals
shown in Fig.\,\ref{fig:HIPresiduals}, left panel. There are several outliers
and fit quality is poor ($\chi^2_{\rm{red}} = 2.01$, RMS(O$-$C)$=
1.43$\,mas). 
We obtain a best fit parallax of $\varpi_{\rm{all}} = 4.37 \pm 0.27$\,mas.

We therefore discarded a total of 6 data points on the basis of their excess
residual from the following satellite orbits: $180$, $252$, $396$, $759$,
$1786$, and $2126$.
We then repeated the 5-parameter fit to $89$ measurements and obtained a better
fit (Fig.\,\ref{fig:HIPresiduals}, right panel; $\chi^2_{\rm{red}} = 0.65$,
RMS(O$-$C) $= 0.81$\,mas), as well as smaller parallax: $\varpi_{\rm{5parm}} =
4.09 \pm 0.16$\,mas.
Note that there is a discrepancy of $\Delta \varpi = 0.38$\,mas
($0.21$\,mag) between the parallax value published in the \citet[$\varpi = 3.77
\pm 0.16$\,mas]{2007ASSL..350.....V} reduction  and our result based on the IAD.
There may be several reasons for such a discrepancy, including:
\begin{enumerate}
\item Our fitting routine. We excluded this possibility by processing other
stars, in particular \DC 's visual companion HD\,213307 for which we obtained
a parallax of $3.69\pm0.46$ mas in exact numerical agreement with the
published value \cite[$3.69\pm0.46$ mas,][]{2007ASSL..350.....V}. As an
additional cross-check, we applied the same methodology to more than $20$
other Cepheids using the same routine and obtained results that agree well with
the published values of \citet{2007ASSL..350.....V}. 
\item our selection of IAD. As noted in the IAD header, $3\%$ of the
IAD were discarded to obtain the solution of \citet{2007ASSL..350.....V},
although it is not specified which ones. We attempted to reproduce their
result by discarding only 3 measurements but did not succeed.
\item the cluster solution. It appears that \citet{2007ASSL..350.....V} used a
special procedure for cluster stars to better remove outlier measurements for
individual stars, under the assumption that all cluster stars are at
indistinguishable distance. While our parallax result is larger
than the previously published value, it is well within the wide range of  
parallaxes ($2.57 - 5.28$\,mas) reported for presumed members of Cep\,OB6.
\item other unknown procedures or modifications that may have been
applied specifically to \DC 's parallax to achieve the solutions presented by
\citet{2007ASSL..350.....V} and \citet{2007MNRAS.379..723V}.
\end{enumerate}

Our parallax estimate ($\varpi = 4.09 \pm 0.16$\,mas) is also considerably
($2\sigma$) larger than the {\it HST}-based result by \citet[$\varpi = 3.66
\pm 0.15$\,mas]{2002AJ....124.1695B}.
Such an increase in parallax would increase \DC 's absolute magnitude by
$0.24$\,mag, making it intrinsically fainter than previously thought
\citep[$M_V = -3.23$\,mag using the absolute magnitude published
by][]{2002AJ....124.1695B}.
We note that the {\it HST}-based period-luminosity relations presented by
\citet{2007AJ....133.1810B} seem to agree better with such an increase in
absolute magnitude for \DC .

The {\it HST} observations were taken at an orbital phase at which
a significant parallax bias due to orbital motion is not very likely, cf.
Fig.\,\ref{fig:rvorbit}.
However, the assumption of a physical association between HD\,213307 as well as
\DC\ in the loose association Cep\,OB6 was required to ``reduce [the {\it HST}]
astrometric residuals to near-typical levels''
\citep[Sec.\,5.5]{2002AJ....124.1695B}.

Concerning \DC 's membership in Cep\,OB6, we note that the association's
discovery did not take into account radial velocity data, since only little such
information was available at the time \citep{1999AJ....117..354D}.
Inspection of the {\tt SIMBAD} database yields radial velocity information for
10 of the 19 presumed member stars (in addition to \DC), with values ranging
from $-7$\,\kms\ to $-38$\,\kms\ and median uncertainty of $1.2$\kms\ based on
measurements by:
\citet{1996BICDS..48...11F,1999A&AS..137..451G,2005A&A...430..165F,2006AstL...32..759G,2007AN....328..889K}.
The RVs are distributed as follows: HIP110807 and HIP112998 have $v_r \sim -7
\pm 2$\,\kms , HIP110497 has $v_r = -13.3 \pm 0.5$\,\kms , HIP109492 and
HIP110988 (= HD213307) are close to \DC 's $v_\gamma = -16.8$\,\kms (to within
$1$\,\kms), HIP113993 has $v_r = -20.9 \pm 1.1$\,\kms , HIP113316 has $v_r =
-25.6 \pm 8$\,\kms, and three stars (HIP110266, HIP110275, HIP110356) have $v_r
< -30$\,\kms\ with reported uncertainties between $3$ and $15$\,\kms .
This relatively wide range of RVs suggests that Cep\,OB6 is not gravitionally
bound\footnote{For this to be the case, the RV dispersion of member stars should
not exceed a few \kms\ \citep[e.g.][]{1986HiA.....7..481M}.}, i.e.
that the assumption of a common distance for all presumed member stars is not valid. This interpretation is corroborated by the wide range of
parallax values of the presumed member stars (see item 3 above). However, more
homogeneous and high-quality radial velocity measurements of the presumed
members of Cep\,OB6 are required to further illuminate this issue.

Although the above suggests that not all of Cep\,OB6's presumed members
have indistinguishable distance, we note that the observational evidence does
support the physical association between HD\,213307 and \DC , since their
parallaxes and radial velocities agree to within the uncertainties. A
re-assessment of the {\it HST} astrometric data without the assumption of
cluster membership (this affects e.g. the spectrophotometric parallax for
HD\,213307, which is part of the {\it HST} reference frame), and taking into
account \DC 's orbital motion would be useful for testing whether the difference
between the published {\it HST} parallax and our result can be reconciled.

\subsubsection{Orbit Analysis}\label{sec:HIPorbit}

We search for an orbital signature in {\it Hipparcos} IAD using the
methodology described in \citet{2011A&A...525A..95S}, which 
has been shown to reliably detect such orbital signatures in {\it Hipparcos}
IAD \citep{2011A&A...525A..95S,2011A&A...528L...8S,2013A&A...556A.145S}.
We use the spectroscopic orbital parameters given in
Tab.\,\ref{tab:orbitsolution} to fit the IAD with a seven-parameter model, which
has the free parameters inclination $i$, longitude of the ascending node
$\Omega$, parallax $\varpi$, and offsets to the coordinates $(\Delta
\alpha,\,\Delta \delta)$ as well as offsets to the proper motions
$(\Delta\mu\alpha,\,\Delta\mu\delta)$.
We then search a two-dimensional grid in $i$ and $\Omega$ for its global
$\chi^2$-minimum with a nonlinear minimization procedure. We determine the
statistical significance of the derived astrometric orbit via a permutation test
\citep{2001ApJ...562..549Z} for which we employ $1000$ pseudo-orbits and derive
parameter uncertainties using Monte Carlo simulations that include propagation
of RV parameter uncertainties.

\begin{figure}
\centering
\includegraphics[scale=0.75]{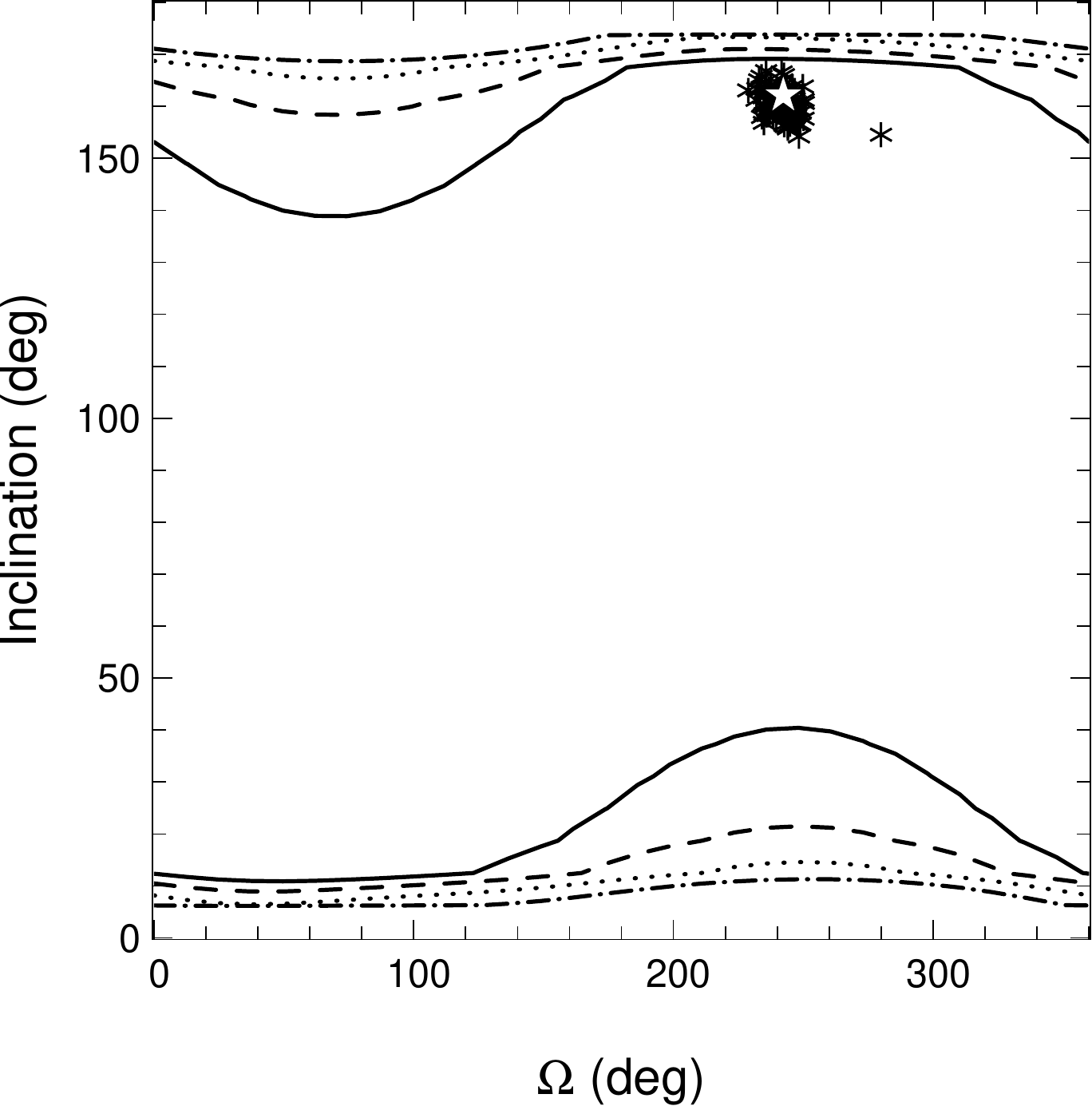}
\caption{Joint confidence contours on the $i-\Omega$-grid. Contour lines
correspond to confidence levels at $1\sigma$ (solid), $2\sigma$ (dashed),
$3\sigma$ (dotted), and $4\sigma$ (dash-dotted). Crosses indicate
the position of the best non-linear adjustment solution for each of the 100
Monte Carlo samples of spectroscopic parameters; the star corresponds to
the adopted parameters.}
\label{fig:HIPiomega}
\end{figure}

Only $41\%$ of the orbit are probed by {\it Hipparcos} measurements, whose
average measurement uncertainty of $1.07$\,mas is furthermore greater than the
barycentric minimum semimajor axis of the orbit ($0.84$\,mas).
It may therefore not surprise that we determine virtually identical results for
5-parameter (single star) and 7-parameter (binary) models, with flat residuals
even for the 5-parameter model (Fig.\,\ref{fig:HIPresiduals}).
From an F-test, we obtain a probability of $4.4\%$ that the single-star model is
true, and the permutation test yields an orbit detection significant at the
$1.8\sigma$ level ($93.7\%$). While the evidence for the orbital signature is
only marginally significant, it has been determined with two independent methods
that are in agreement.
It thus appears that there is some temporal coherence present in the {\it
Hipparcos data}, although we cannot claim orbit detection using {\it Hipparcos}
astrometry.
We can, however, use our orbital analysis to set (very) loose constraints on the
inclination ($10^\circ \lesssim i \lesssim 170^\circ$) of the orbit, see
Fig.\,\ref{fig:HIPiomega}, allowing us to place an upper limit on the companion
mass with $0.2$\,\Msol\ $\lesssim M_2 \lesssim 1.2$\,\Msol, which is fully
consistent with the lack of spectral features due to the companion in the {\it
Hermes} data set, cf. Sect.\,\ref{sec:spectroscopy}.

While we are unable to claim detection of the orbit from {\it Hipparcos}
astrometric data, we caution that the previous estimate of \DC 's proper motion
may have been affected by the companion.

\subsection{Gaia}\label{sec:Gaia}
The {\it ESA} space mission {\it
Gaia}\footnote{\url{http://www.cosmos.esa.int/web/gaia}} is currently conducting
an unprecedented census of our Galaxy, measuring position, proper motion, and
parallax for more than a billion stars during a nominal mission duration
of five years.
\DC\ will be among these billion objects thanks to {\it Gaia}'s ability to
observe very bright stars\footnote{Stars brighter than $G$-band
magnitude 5.7 will be heavily saturated and need special treatment to determine their
centroids. We conservatively assume a single observation
per {\it Gaia} field-of-view transit.} \citep{2014SPIE.9143E..0YM}.
For an assumed single-measurement precision of $\sigma_{\rm{Gaia}} \approx
100\,\mu$as, the minimum barycentric semimajor axis ($a_1 \sin{i} \approx
840\,\mu$as (Sect.\,\ref{sec:orbit})) is roughly $8$ times larger than the
measurement uncertainty.
Figure\,\ref{fig:rvorbit} shows the range of orbital phases covered during the
nominal mission duration plus a possible one year extension. It shows that {\it
Gaia} measurements will cover the majority of the orbit and, crucially, will be
measuring astrometry during periastron passage. We investigate the detectability
of \DC 's orbit from {\it Gaia} astrometry in two ways.

First, the detectability of astrometric orbits with {\it Gaia}
can be estimated from a consideration of the astrometric signal-to-noise $S/N =
a_1 \, \sqrt{N} / \sigma_{\rm{Gaia}}$
\citep{2011A&A...528L...8S,2015MNRAS.447..287S}.
Assuming $86$ {\it Gaia} observations of \DC\ during the 5-year mission (see
below) and accounting for $10\%$ dead time with an assumed $100\,\mu$as accuracy
for individual measurements and the minimum amplitude of $\sim\! 840\,\mu$as, we
obtain $S/N \approx 73$, which is much higher than the detection threshold of 20
described in \citet{2015MNRAS.447..287S}. This indicates that {\it Gaia} will
detect the astrometric orbit of \DC , despite not 
measuring at all orbital phases.

Second, we use the Gaia astrometric simulation software {\tt AGISlab}
\citep{2012A&A...543A..15H} to investigate the detectability of \DC 's orbit
derived in Sect.\,\ref{sec:orbit}. We adopt the $\Delta\chi^2$ metric from
\citet{2014ApJ...797...14P} to evaluate the detectability of the
astrometric binary signal. This metric measures the reduction in minimum
$\chi^2$ when going from a (single-star) 5-parameter solution to a (binary) 12-parameter
Keplerian solution. Even for our adopted worst-case scenario\footnote{This
assumes the worst possible configuration of sky-alignment and observation noise,
as well as $20\%$ dead time}, we find an improvement of 
$\Delta\chi^2_{\rm{min,3\sigma}} = 139$ when accounting for the orbital motion.
This is well above the threshold for precise (parameters determined to better
than $10\%$) orbit characterization \citep[$\Delta\chi^2 > 100$]{2014ApJ...797...14P}.
We therefore predict that {\it Gaia} will clearly detect and characterize the
astrometric counterpart to the spectroscopic orbit discovered here.

\section{Discussion: Piecing Together the Puzzle}\label{sec:puzzle}
Now that \DC 's nature as a spectroscopic binary is revealed, it is worth
revisiting other observed features of the prototype of classical
Cepheids in this new light. 

\citet{2014ApJ...794...80E} recently provided evidence that \DC\ is a soft X-ray
source with a luminosity of $L_X (0.3 - 2 \rm{keV}) \approx 4.5 - 13 \times
10^{28}\,\rm{erg\,s^{-1}}$ and peak flux at $kT = 0.6 - 0.9$\,keV. As these
authors discuss, young \citep[$\sim 120$\,Myr for a $5$\,\Msol\ Cepheid,
according to Geneva evolution models
by][]{2012A&A...537A.146E,2013A&A...553A..24G} low-mass main-sequence companions
can provide coronal X-ray emission. In Sects.\,\ref{sec:orbit} and
\ref{sec:HIPorbit}, we constrained the mass range for the unseen companion to be
 $0.2 < M_2 < 1.2$\,\Msol\ based on radial
velocities and {\it Hipparcos} astrometric measurements. As Cepheids are
rarely detected in X-rays, it appears likely that \DC 's young
main sequence companion is responsible for the detected variable X-ray
activity.

Due to the high eccentricity of the orbit, \DC\ and its companion have recurrent
close encounters, with a pericenter distance of $r_{\rm{per}} = (1 -
e)a_{\rm{rel}} = 1.89$\,au\,$= 409$\,\Rsol\ $= 9.5\,R_{\star}$, where $R_\star =
43.3$\,\Rsol\
\citep{1988AJ.....96.1565T,2005A&A...438L...9M,2008ApJ...674L..93N}.
It is also important to bear in mind that \DC\ is currently in the core He
burning phase, most likely on the second crossing of the instability strip as
shown by the decrease in pulsation period (measured e.g. by E.\,Hertzsprung and
reported by \citealt{1919Obs....42..338E},
\citealt{2000ASPC..203..244B}, and \citealt{2014ApJ...794...80E}).
It has therefore previously occupied the red giant branch, where its radius was
even larger, approximately $R_{\star,RG} \sim 80$\,\Rsol .
Assuming an unchanged orbit, pericenter passage would have brought the two stars
to within $5 \times R_{\star,RG}$.

Taking these considerations further, we consider the possibility of previous
interactions between \DC\ and its companion. Since \DC\ is on a highly eccentric
orbit and the Roche-lobe formalism is valid only for circular orbits, we adopt
the Roche-lobe formalism in the quasi-static approximation as presented by
\citet{2007ApJ...660.1624S}.
Assuming $M_{\delta\,Cep} = 5.2$\,\Msol , $M_{2} = 0.7$\,\Msol , and equatorial
velocity $v_{eq} = 10$\,\kms , we obtain the volume-equivalent Roche lobe
radius at pericenter of $R_{\rm{Roche,corr}} = 1.05$\,au $\approx
5.2\,R_{\star} \approx 2.8\,R_{\star,RG}$.
Hence, at pericenter passage \DC\ fills $19\%$ of its (quasi-static) Roche lobe,
and the Roche lobe radius at pericenter is approximately $55\%$ of the distance
at pericenter passage ($r_{\rm{per}} = 1.89$\,au).
At apocenter, the distance between the two stars is a factor $(1+e)/(1-e) = 5.1$
larger, and \DC\ fills only about $3.7\%$ of its Roche lobe.
\DC\ may thus become noticeably deformed due to tidal interactions close
to pericenter passage, while being spherical near apocenter passage.
During the red giant phase, this situation would have been even more extreme, if
the past orbit was similar to the present-day orbit.
The above considerations suggest that the observed circumstellar material
\citep{2010ApJ...725.2392M,2006A&A...453..155M} and bow-shock
\citep{2012ApJ...744...53M} may originate from previous and ongoing binary
interactions. A more detailed investigation of such interactions is required to
discuss this scenario in terms of \DC 's mass loss history \citep[see the
discussion in][]{2012ApJ...744...53M}, but is considered out of scope for this
paper.

If \DC\ and its companion have a history of episodal interactions at pericenter
passage, then the high eccentricity of the orbit may appear surprising.
However, \citet{2007ApJ...667.1170S} showed that rapid circularization is not
expected for all close-in binaries with this type of interactions. Applying our
results for \DC\ to their formalism shows that \DC 's orbit is not expected
to have been rapidly circularized.
Furthermore, \DC\ is a visual binary whose outer companion is understood to be
physically associated and on a very long-period orbit
\citep{2002AJ....124.1695B}. The high eccentricity of the inner binary (the one
shown in the present work) could thus have been driven up by the outer companion
by a Kozai-Lidov mechanism \citep{1962P&SS....9..719L,1962AJ.....67..591K} 
and may have varied significantly over its evolutionary history.

\begin{figure}
\centering
\includegraphics[scale=0.9]{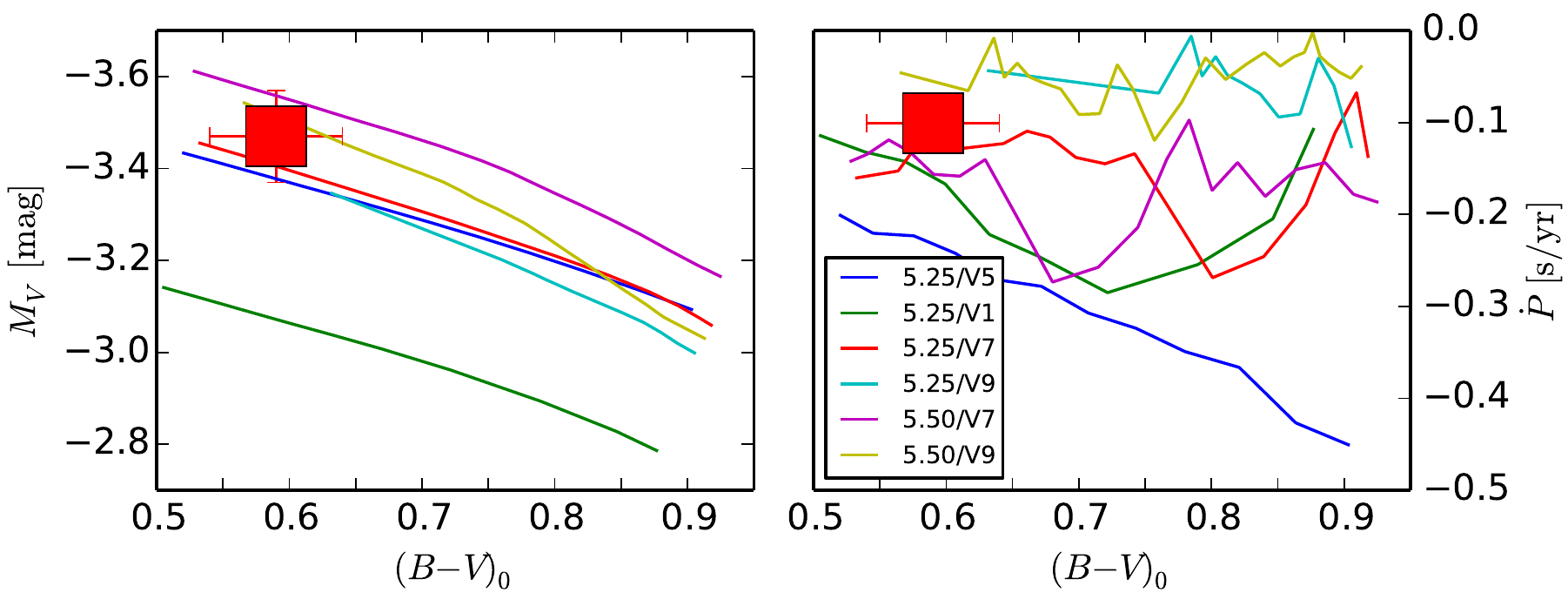}
\caption{\DC 's color-magnitude and color-rate of period change
diagrams that compare the measured quantities with predictions from Solar
metallicity Geneva stellar evolution models
\citep{2012A&A...537A.146E,2013A&A...553A..24G,2014A&A...564A.100A}.
Note that the predicted rates of period change are based on the evolution of the
average density, i.e., they have not been determined from a pulsation
code.
We use absolute magnitude and dereddened color by \citet{2002AJ....124.1695B} and the
rate of period change measured by \citet{2014ApJ...794...80E} as the observed values. The left panel
is particularly well-suited for determining progenitor mass, whereas the rate of
period change is very sensitive to the main sequence rotation of \DC 's
progenitor. The legend in the right panel indicates ZAMS mass in solar units
followed by the model's initial rotation rate, where V1 corresponds to $\omega
= \Omega / \Omega_{\rm{crit}} = 0.1$, V7 to $\omega=0.7$, and V9 to $\omega =
0.9$. V5 corresponds to $\omega = 0.568$, i.e., $v / v_{\rm{crit}} = 0.4$, see
\citet{2012A&A...537A.146E}. Comparing the observed values to model predictions
(assume single star evolution), \DC\ is consistent with a progenitor mass of
approximately $5.25$\,\Msol\ and slightly faster-than average  surface rotation
($\omega \sim 0.7$). Adopting our new parallax estimate ($\varpi = 4.09
\pm 0.16$) would change $M_V$ to $-3.23$\,mag, resulting in a lower inferred
mass of $5.0$\,\Msol\ while yielding the same result in terms of rotation}.
\label{fig:dCepPdotCMD}
\end{figure}

If \DC\ and its companion have undergone significant interactions, one might
expect the evolutionary status of \DC\ to vary significantly from that of a
Cepheid with a single star progenitor. We therefore examine \DC 's current
evolutionary status in Figure\,\ref{fig:dCepPdotCMD} to search for signs of
non-standard evolution. We compare observed absolute $V$ magnitude, $(B-V)_0$
color \citep[both from][]{2002AJ....124.1695B}, and rate of (pulsation)
period change $\dot{P_{\rm{puls}}} =  -0.1006 \pm 0.0002$
\citep{2014ApJ...794...80E} with predictions from Geneva stellar
evolution models\footnote{Models accessible at
\url{http://obswww.unige.ch/Recherche/evoldb/index/Interpolation/}} of
Solar metallicity that include rotation
\citep{2012A&A...537A.146E,2013A&A...553A..24G} and have been studied
specifically in the context of classical Cepheids by
\citet{2014A&A...564A.100A}. Strong disagreement between the observations and
these predictions could be considered evidence for binary interactions, since no
binary interactions are accounted for in these model predictions.
However, we find that \DC 's location in both diagrams is consistent with a
$5.25$\,\Msol\ Cepheid whose progenitor had a slightly faster-than-average
initial rotation ($\Omega / \Omega_{\rm{crit}} \sim 0.7$).
Assuming these parameters for the evolutionary models yields an age of
$112$\,Myr.
Adopting our new parallax from Sec.\,\ref{sec:HIPpar} yields a
smaller initial mass of $M \approx 5.0$\,\Msol\ and thus an older age of $127$\,Myr,
while leaving the implications regarding the rotational history unchanged.

At the present level of accuracy, we thus do not 
find any irreconcilable discrepancies 
between the predicted and observed evolutionary states of \DC . This shows that
binary interactions, if present, have either had a negligible effect on the
evolutionary path of \DC , or that the interactions are weak and
slow enough for \DC\ to reach an equilibrium state similar to a non-interacting
star.

In summary, the discovery of \DC 's nature as a spectroscopic binary
helps to complete the puzzle created by a plethora of observations. While there
is no clear evidence for a non-standard evolutionary path, \DC\ is a
particularly interesting example of the limitations that this evolutionary phase
is subject to for binary stars \citep[cf.][]{2014arXiv1412.3468N} and deserves
detailed observational follow up and dynamical modeling to further
investigate its intriguing past that may have been marked by tidal interactions
due to both the inner (discovered here) and outer (HD\,213307) companions.

\section{Conclusions}\label{sec:conclusions}
$230$ years after the discovery of its variability
\citep{1786RSPT...76...48G}, we discover the spectroscopic binary nature of \DC
, archetype of classical Cepheid variable stars and one of the most-studied variable stars.

Our discovery is demonstrated using new high-precision
radial velocities measured from high-quality optical spectra obtained with the high-resolution
spectrograph \hermes .
Combining these new high-precision data with lower-precision RVs from the literature, we
determine the orbital solution for the spectroscopic binary, which is an inner
binary to the outer visual binary system discussed by
\citet{2002AJ....124.1695B}. \DC\ thus appears to be a pair of binary
stars.

We re-analyze {\it Hipparcos} intermediate astrometric data (cf.
Sects.\,\ref{sec:HIPpar}) and obtain the parallax $\varpi = 4.09 \pm
0.16$\,mas ($d = 244 \pm 10$\,pc) using a 5-parameter (single star) model.
This result is larger than the estimates reported by \citet[{\bf $\varpi =
3.77\pm 0.16$\,mas}]{2007ASSL..350.....V}, \citet[$\varpi = 3.71 \pm
0.12$\,mas]{2007MNRAS.379..723V}, and \citet[$3.66 \pm
0.15$\,mas]{2002AJ....124.1695B}. While these previously published results based
on {\it Hipparcos} and {\it HST} astrometry agree
to within their stated uncertainties, they all shared a common assumption of
\DC 's membership in the loose association Cep\,OB6, which our analysis does
not and which we argue should be revisited using radial velocities. 
Relaxing this assumption and accounting for orbital motion in a re-analysis of
the {\it HST} astrometric data would  be useful to test whether the
existing {\it HST} astrometry can be reconciled with our {\it Hipparcos-}based
result.

We perform an orbital analysis of the {\it Hipparcos} IAD in
Sect.\,\ref{sec:HIPorbit} and find tentative evidence for an orbital signature, although no detection can be claimed.
Based on detailed simulations, we show that {\it Gaia} is highly sensitive to
the astrometric orbit of \DC\ and will likely model the full set of Keplerian
parameters with better than $10\%$ accuracy. The orbit will have to be accounted
for when determining proper motions from {\it Hipparcos} and {\it Gaia}
data.
 
Using the constraints provided by the optical spectra, the orbit measured from
RVs, the astrometric orbital analysis, and assuming a mass of $5.0 -
5.25$\,\Msol , we constrain the mass range of the companion to be $0.2 <
M_2 < 1.2$\,\Msol . Adopting the lower
mass for \DC\ mainly affects the upper mass limit, which would become
$1.1$\,\Msol\ in this case. Given that the spectroscopic companion is
expected to be the same age as \DC , i.e., approximately $100-130$\,Myr (depending on mass and ZAMS rotation
rate of the progenitor), the reported X-ray emission detected using {\it
XMM-Newton} \citep{2014ApJ...794...80E} could be explained by magneto-rotational
activity of a young main-sequence star.

The close periastron approach of the two stars has potentially far-reaching
consequences for the explanation of the observed circumstellar environment of
\DC . Detailed modeling of the orbital and stellar evolution of this complex
system is desirable to further improve our understanding of the archetype of
classical Cepheids and its intriguing past.

\acknowledgments
 
  Many thanks are due to everyone who aided in securing the analyzed 
  datasets and, in particular, to the \hermes\  and {\it Mercator} teams.
  Damien S\'egransan is acknowledged for assistance regarding the use of {\tt
  Yorbit}. RIA acknowledges Miranda A. Gaanderse for her creative assistance
  in finding the title for this paper and Dr. Paul I. Anderson for his
  careful reading of the manuscript. We thank the anonymous referees for
  their constructive reports.
  
  This research is based on observations made with the Mercator Telescope,
  operated on the island of La Palma by the Flemish Community, at the Spanish Observatorio del Roque de
  los Muchachos of the Instituto de Astrofísica de Canarias.
  \hermes\ is supported by the Fund for Scientific Research of Flanders (FWO),
  Belgium, the Research Council of K.U.~Leuven, Belgium, the Fonds National de
  la Recherche Scientifique (F.R.S.-FNRS), Belgium, the Royal Observatory of Belgium, the
  Observatoire de Gen\`eve, Switzerland, and the Th\"uringer Landessternwarte,
  Tautenburg, Germany. This study also employed spectral data
  retrieved from the {\it ELODIE} archive at Observatoire de Haute-Provence (OHP).
  
  This research has made use of NASA's ADS Bibliographic Services, the SIMBAD
  database and the VizieR catalogue access tool provided
  by CDS, Strasbourg, and Astropy, a community-developed core Python package for Astronomy
  \citep{2013A&A...558A..33A}.
   
  RIA acknowledges funding from the Swiss NSF. JS is supported by an ESA
  research fellowship in space science.

\bibliographystyle{aa} 
\bibliography{Bib_Berry,Bib_mine,Bib_Cepheids,Bib_interferometry,Bib_DistanceScale,Bib_Spectroscopy,ThesisBibTexRefs,RotatingCepBibTexRefs,Bib_StellarEvolution,Bib_Spectropolarimetry,Bib_general}

{\it Facilities:} \facility{Mercator1.2m}, \facility{HIPPARCOS}

\end{document}